\documentclass[twocolumn]{aastex63}
\hypersetup{linkcolor=red,citecolor=green,filecolor=cyan,urlcolor=magenta}

\newcommand\Swift{\emph{Swift} }
\newcommand\Fermi{\emph{Fermi} }
\title{Classifying blazar candidates from 3FGL unassociated catalog into BL Lacs and FSRQs using WISE and \Swift$-$XRT}
\shorttitle{Kaur et al. 2020}
\shortauthors{Kaur et al.}
\submitjournal{AJ}

\begin{document}
\title{Classifying blazar candidates from the 3FGL unassociated catalog into BL Lacs and FSRQs using \Swift and WISE data}

\author[0000-0002-0878-1193]{Amanpreet Kaur}
\affiliation{Department of Astronomy and Astrophysics \\
 Pennsylvania State University
University Park, PA 16802, USA}

\author[0000-0002-5068-7344]{Abraham D. Falcone}
\affiliation{Department of Astronomy and Astrophysics \\
 Pennsylvania State University
University Park, PA 16802, USA}

\author[0000-0002-3019-4577]{Michael C. Stroh}
\affil{Center for Interdisciplinary Exploration and Research in Astrophysics (CIERA), Northwestern University, Evanston, IL 60201, USA}

\begin{abstract}
We utilize machine learning methods to distinguish BL Lacertae objects (BL Lac) from Flat Spectrum Radio Quasars (FSRQ) within a sample of likely X-ray blazar counterparts to Fermi 3FGL unassociated gamma-ray sources. From our previous work, we have extracted  84 sources that were classified as $\geq$ 99\% likley to be blazars.  We then utilize Swift$-$XRT, Fermi, and WISE (The Wide-field Infrared Survey Explorer) data together to distinguish the specific type of blazar, FSRQs or BL Lacs. Various X-ray and Gamma-ray parameters can be used to differentiate between these subclasses. These are also known to occupy different parameter space on the WISE color-color diagram. Using all these data together would provide more robust results for the classified sources. We utilized a Random Forest Classifier to calculate the probability for each blazar to be associated with a BL Lac or an FSRQ. Based on P$_{bll}$, which is the probability for each source to be a BL Lac, we placed our sources into five different categories based on this value as follows; P$_{bll}$ $\geq$ 99\%: highly likely BL Lac, P$_{bll}$ $\geq$ 90\%: likely BL Lac, P$_{bll}$ $\leq$ 1\%: highly likely FSRQ, P$_{bll}$ $\leq$ 10\%: likely FSRQ, and  90\% $<$ P$_{bll}$ $<$ 10\%: ambiguous. Our results categorize the 84 blazar candidates as 50 likely BL Lacs and the rest 34 being ambiguous. A small subset of these sources have been listed as associated sources in the most recent Fermi catalog, 4FGL, and in these cases our results are in agreement on the classification.
\end{abstract}


\section{Introduction} \label{sec:intro}
Blazars are a subclass of the Active Galactic Nuclei which have their jets pointing along our line of sight \citep{Blandford1978}. They are further divided into two categories; Flat Spectrum Radio Quasars (FSRQs) and BL Lacertae Objects (BL Lacs) based on their optical spectra. The FSRQs display broad emission lines, whereas the BL Lacs display no lines or narrow lines (equivalent width $<$ 5 \AA). The spectral energy distribution of these objects displays a characteristic double-bump structure. The lowest energy bump is typically attributed to the synchrotrom emission (radio to X-ray) from electrons propagating in the jet, and the higher energy bump (X-ray to gamma-rays) is typically attributed to the synchrotron self Compton mechanism \citep{Maraschi1994}, and/or external inverse Compton processes, and/or hadronic processes such as proton synchrotron. Understanding the origin of the differences between the two subclasses of blazars has been one of the open questions in the field. The idea of the blazar sequence, termed by \citet{Fossati1998} and re-visited by \citet{Ghisellini2008, Ghisellini2017}, revealed that the most luminous blazars possess lower synchrotron frequencies, which often is the case in FSRQs, and vice versa for BL Lacs. The blazar sequence has been challenged by finding highly-luminous, high-synchrotron-peaked blazars, e.g. \citet{Padovani2012}. Moreover, \citet{Ghisellini1998} suggested a unified scheme for blazars in which FSRQs eventually evolve into BL Lacs once their accretion disk is exhausted. The observational evidence of finding FSRQs at typically higher redshifts than that of BL Lacs suggests this scenario. However, several (29) BL Lacs have been found at high redshifts e.g., \citep{Rau2012, Kaur2017, Kaur2018, Rajagopal2020}. These blazar parameter space and the theories regarding their evolution need to be further explored by obtaining a more complete sample of both types of blazars. The Fermi Gamma Ray Observatory has revealed more than 5000 sources since its launch in 2008. Blazars constitute the bulk of the overall known extragalactic gamma-ray population ($>$ 75\%) in all the Fermi catalogs; 1FGL \citep{Abdo2010a}, 2FGL\citep{Nolan2012}, 3FGL \citep{Acero2015}, and 4FGL\citep{Collaboration2019}. The galactic population is dominated by pulsars ($\sim$ 8\% of the total gamma-ray population). However, each catalog represents approximately one-third of its sources as unassociated or unknown. Finding associations to these sources or classifying these is a multi year task which often requires multiwavelength observations for confirmations. In the past few years, various studies have been conducted on these gamma-ray sources where machine learning methods were employed to classify the unassociated sources; e.g., \citep{Ackermann2012a,Parkinson2016,Marchesini2020}. In particular, \citet{Marchesini2020a,Marchesini2020} first explored the connection between gamma-rays and X-rays for blazars and later utilized X-ray data from \Swift in conjunction with the \Fermi gamma-ray data to find BL Lacs among the unassociated sources.. We \citep{ Falcone2015, Kaur2019b} have conducted an X-ray survey targeting the Fermi unassociated source fields and found various possible X-ray associations\footnote{www.swift.psu.edu/unassociated}. Since the majority of the known gamma-ray sources are blazars and pulsars, it is highly likely that a rather large population of the unassociated sources could belong to these two populations. It should be noted that the sensitivity for \Swift-XRT is $\sim$ 1.0 $\times$ e$^{-13}$ erg/cm$^{2}$/s for a 4 ksec exposure, which was the average exposure for this survey. Most of the known \Fermi blazars, as well as some of the known Fermi pulsars, are detectable with high signal-to-noise ratio within this exposure time, as shown in Fig. 1-4 in \citet{Kaur2019b}. These authors utilized machine learning methods on these X-ray counterparts to find blazars and pulsars. Their results yielded 134 blazars and 8 pulsars with high probabilities based on the machine learning methods.\\
In this work, the objective is to classify the highly probable blazar candidates revealed from our previous work into subclassifications of BL Lacs and FSRQs using the methods of Machine Learning. The paper is divided as follows. Section~\ref{sec:sample} describes the process of the final sample selection and Section~\ref{sec:Analysis} explains the overall analysis method. The results and the conclusions of the this study are published in Sections ~\ref{sec:results} and \ref{sec:conclusions}, respectively. 

\section{Sample Selection} \label{sec:sample}
The \Swift-Xray Telescope (XRT) \citep{Burrows2005} conducted observations for 803 3FGL unassociated sources in order to search for their X-ray counterparts. These were chosen at random from the complete sample of $\sim$ 1500 unassociated sources provided they were (i) not listed as a confused source in the 3FGL catalog, (ii) not listed as extended in the 3FGL catalog, and (iii) the 3FGL source 95\% uncertainty region has a semi-major axis that is smaller than 10 arcmin, thus enabling it to fit within the \Swift field of view. All these X-ray observations were completed through a \Swift fill-in program, with an average XRT exposure time for each 3FGL field of approximately 4 ksec, and the results have been provided online at https://www.swift.psu.edu/unassociated/.  It is possible that a small number of X-ray sources found during this campaign are actually spurious associations with the 3FGL source. This probability can be estimated based on the known \Swift-XRT sensitivity for detecting an X-ray source during the randomly distributed \Swift-XRT exposures on the 3FGL fields (the 3FGL follow-up fields were distributed across the whole sky and randomly chosen for \Swift-XRT follow-up as part of a fill-in observation program) coupled with the LAT error ellipse for the sources in our sample. We found this sensitivity by using 'empty' \Swift-XRT fields distributed across the sky, using GRB fields with the GRB masked out, and calculating the spurious source detection density in these fields as a function of exposure time, using the same source detection criteria that is utilized for finding possible X-ray counterparts to Fermi unassociated sources. The majority of the \Swift-XRT follow-up observations of the \Fermi unassociated sources in our sample were of roughly 4 ksec exposure time.  To be included in our sample, a potential X-ray counterpart had to be detected at the $>$4$\sigma$ signal-to-noise threshold.  For a 4 ksec \Swift-XRT exposure, less than $\sim$1 random X-ray detection is expected for every 100 Fermi-LAT error ellipses, when using the 4$\sigma$ threshold that we used for X-ray source selection and when using a typical \Fermi-LAT 95\% confidence ellipse for an unassociated gamma-ray source.  However, a subset of our observations received significantly longer exposures and some of the Fermi-LAT error ellipses are larger than typical, thus increasing the chance coincidence probability in those cases.  By using the actual \Swift-XRT exposure and the Fermi-LAT error ellipse of each follow-up field, we found that the median chance probability of a spurious 4$\sigma$ X-ray source detection was less than 0.01; aside from a few outliers, there was generally a low chance that the given X-ray source counterpart candidate is not actually associated with the gamma-ray source. These chance probabilities are tabulated below, along with the results of this work. Among these, 217 X-ray sources were found which met the following criteria: (1)  only one X-ray source within the 95\% 3FGL uncertainity region, (2) this possible X-ray counterpart is detected at a signal to noise ratio $\ge$ 4. \citet[hereafter, K19]{Kaur2019b}
performed a machine learning analysis (random forest) to find pulsar and blazar candidates among these 217 X-ray counterparts to the 3FGL unassociated sources. According to the random forest classifier method used in K19, 134 sources from the sample resulted in classifications as highly likely blazars, i.e. the sources for which the probabilities to be associated with the blazar class was $\ge$99\%. See Section 3 for the random forest analysis method and Section 4 for the classification criteria in K19. For this work, we selected these 134 sources which were found to be highly likely blazar candidates. Since the time of the K19 publication, a few more \Fermi unassociated sources were observed with \Swift$-$XRT, and 25 of these were found to have exactly one source within the \Fermi uncertainty circle with SNR $\geq$ 4. We applied the K19 criteria and ML methods to these 25 X-ray sources in order to form a more complete list of blazar candidates for this work. P$_{bzr}$ was defined as the probability for a source to be a blazar, which were yielded by the RF classifier. As was done in K19, the sources were classified into one of the following categories: pulsar (P$_{bzr}$ $\le$1\%, likely pulsar (P$_{bzr}$ $\le$ 10\%), blazar (P$_{bzr}$ $\ge$ 99\%), likely blazar (P$_{bzr}$ $\ge$ 90\%) and ambiguous (90\% $\le$ P$_{bzr}$ $\ge$ 10\%).  The results of the 25 new sources are presented in Table~\ref{tab:new sources}. Among these, 7 were found to be highly likely blazar candidates. These 7 sources were added to our initial 134 source sample of likely X-ray blazar counterparts to 3FGL unassociated sources. Overall, we obtained 139 sources which are highly likely blazars. The \Swift-XRT positions of these sources were utilized to search for any positional associations in the AllWISE catalog \citep{Cutri2013} within 5 arcsec positional uncertainty. An average uncertainty associated with an XRT position for these data is less than $\sim$ 5''; therefore, only if a WISE source was found within this search radius, it was assumed to be the WISE counterpart to this source. While it is possible that a small number of these assumed WISE counterparts could be spatially coincident by random chance, this is likely to be the case for only a few of the WISE sources. 
We randomly selected sources from the \Swift-XRT blazar catalog which comprises of 2831 blazars using 15 years of data \citep{Giommi2019}. We searched for WISE sources corresponding to these blazar positions within the 5" uncertainity region which is the average \Swift-XRT positional uncertainty. Aside from the WISE counterparts to these blazars, we found that a secondary source within 5" was found for 22 positions which suggests that there is a $\sim$ 0.7\% probability that a WISE source can be found on a random location in the sky within a 5'' uncertainty region. Of course, some of the WISE positions fall within a much smaller radial distance from the XRT positional centroid and WISE counterparts are expected for many XRT sources so this 10.8\% estimate is an upper limit that simply tells us that most of the WISE sources are indeed likely counterparts to the XRT sources. This AllWISE catalog was generated using the Wide-field Infrared Survey Explorer \citet[][WISE]{Wright2010} which provides the fluxes, proper motion, accurate positions for approximately 800 million objects. We found matches for 84 sources in our highly likely blazar candidate sample. We proceeded with further analysis for this final sample of 84 sources as described in the next section.
\section{Analysis}\label{sec:Analysis}
\citet{Massaro2012} introduced a method to find blazars among other sources using WISE colors; W1, W2, W3 and W4 corresponding to 3.4, 4.6, 12 and 22 $\mu$m, respectively. These authors showed that blazars occupy a particular region on a color-color plot (W1-W2 vs W2-W3) in the IR regime (WISE in this case) which separates them from other source types. They termed this region as the WISE Blazar Strip (WBS). Moreover, the two classes of blazars; FSRQs and BL Lacs occupy different regions within this strip. Therefore, these color indices could be utilized to separate one blazar class from another, which is the immediate objective of this work. See Fig.~1,2 in \cite{Massaro2012} for details. Fig.~\ref{fig:wise_all} shows the blazar strip for \Fermi blazars along with the 84 unassociated sources. It is quite apparent from this figure that these are highly likely blazars (also predicted by our ML methods in K19), as they follow the pattern of the known blazars. In addition, the fact that the BL Lacs and FSRQs occupy different parameter space on this color-color plot is clearly shown. In our previous work, we used both X-ray and gamma-ray properties of the \Fermi unassociated sources to distinguish pulsars from blazars using the random forest classifier machine learning algorithm \citep{Breiman2001}. Here we employ the same method, with additional WISE parameters, to the subsample of highly likely blazar candidates which are 84 in number, as described in Section~\ref{sec:sample}. While some distinction between the two classes of blazars can be seen when the five considered X-ray and gamma-ray properties are compared (e.g. X-ray flux, gamma-ray flux, gamma-ray variability index, gamma-ray spectral index, and curvature), the addition of WISE color parameters is expected to enhance this distinction. In this work, we utilize these five X-ray and gamma-ray parameters along with two WISE color indices; W1-W2 and W2-W3.  We compare them simultaneously using the random forest classifier. In order to proceed with this algorithm, a sample that includes both known classes of blazars was required, with each of the blazars in this sample having known values for all of these above mentioned seven parameters.
\subsection{Training and Test data}
\begin{figure}
     \centering
     \includegraphics[width=0.5\textwidth]{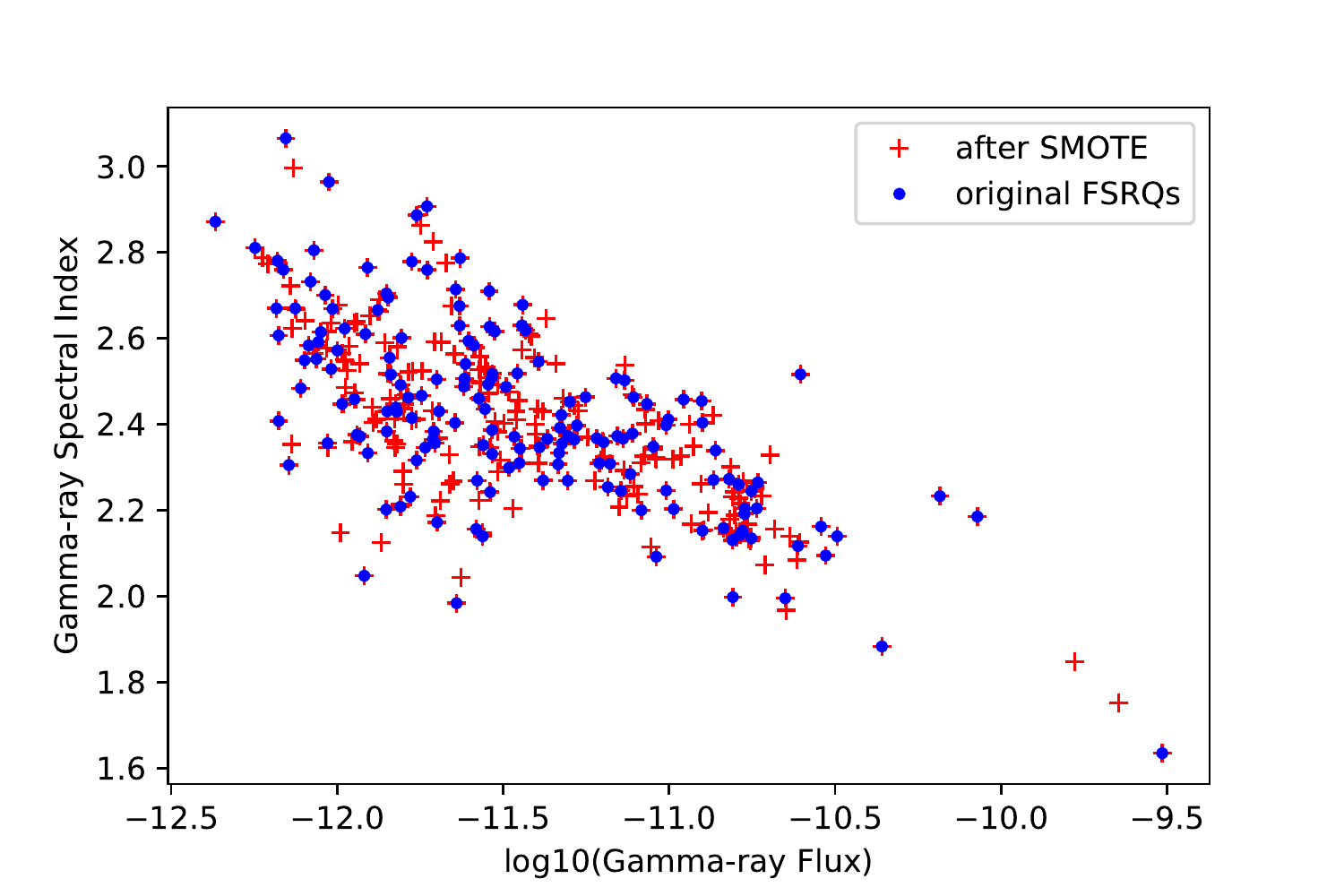}
     \caption{An example two dimensional plot displaying the distribution of the original FSRQs (blue dots) and artificially created FSRQs (red pluses) using SMOTE. The x-axis represents the logarithm of gamma-ray flux (0.1-100) (GeV)[erg/cm$^{2}$/s] and the y-axis displays the gamma-ray spectral index. These parameters were obtained from the 3FGL catalog.}
     \label{fig:smote}
 \end{figure}

A total of 501 known blazars were extracted from \citet[3LAC,][]{Ackermann2015b} for which gamma-ray, X-ray and WISE data were available. The gamma-ray and X-ray properties of these known blazars were obtained from \citet[3FGL, ][]{Acero2015} and \citet[3LAC,][]{Ackermann2015b}, respectively. Fig ~\ref{fig:pairplot} displays the comparison of two subclasses of blazars along with the unassociated sources. It should be noted that the X-ray fluxes for blazars in 3LAC catalog were extracted from the RASS survey \citep{Voges1999,Voges2000}. These flux values are provided in the energy range 0.1-2.4 keV. For the 84 unassociated sources, the X-ray fluxes were derived using \Swift-XRT in the energy range from 0.1 to 2.4 keV for the consistency. The WISE color indices were obtained from the AllWISE catalog. Out of these 501 blazars, 162 were FSRQs and 339 were BL Lacs. The unbalanced proportion of these two classes could lead to biased results towards one particular class, therefore this was corrected by using a class balancing algorithm, SMOTE \citep{Chawla2002}. SMOTE uses the k nearest neighbors method to synthetically generate sources for the underrepresented class to match it with the number of sources in the other class. Here we employed the SMOTE algorithm provided in the \texttt{scikit-learn} library of python which utilizes 5 nearest neighbors to create one synthetic data point. In this case, since FSRQs were 162 by number as compared to 339 BL Lacs. SMOTE algorithm added 177 synthetic data points mimicking the properties of the original FSRQs. This led to our final sample of 339 BL Lacs and 339 FSRQs. An example displaying the results of SMOTE analysis are shown in Fig.~\ref{fig:smote}. Here these results are shown for gamma-ray flux vs the spectral index, but these FSRQs mimic the real FSRQs for all the parameters which would be used to train the classifier. 
\subsubsection{Random Forest Parameter Selection and Accuracy Calculation Method}
We employed the random forest classifier from \texttt{sklearn} using \texttt{python 3.6}, which is a supervised method of machine learning based on the method of decision trees \citep{Breiman2001}. 
A complete description of this method and the details of its implementation are described in section 3.1 in K19. A parameter tuning algorithm \texttt{GridSearchCV} in \texttt{sklearn} was employed to find the optimum parameters for the random forest classifier. Based on this optimization, 1000 decision trees with a maximum depth of 10 splits(nodes) in each tree were employed to obtain the final classification for each source in this method.\\
Generally in a machine learning algorithm, a majority of the complete sample is reserved for the training set and a smaller subset is utilized as the test sample to check the accuracy of the underlying classifier. However, the accuracy obtained from this method is clearly biased since it is based on one given test sample. Therefore, in this work, we employed a 10-fold cross-validation method using \texttt{sklearn} which divided the original sample into 10 equal size subsamples such that one out of these 10 samples was chosen as a test sample (one at a time) and the rest combined were considered a training sample. The trained classifier was then applied to the given test subsample to obtain the accuracy value. This procedure was repeated 10 times to obtain accuracies from each test sample. The overall accuracy was calculated as an average of accuracies obtained from these iterations. This accuracy calculation method has the advantage that it iterates through the complete sample which results in less sample bias, relative to calculating accuracy based on a single test sample. Based on the procedure explained above, the random forest classifier was trained and then cross validated which yielded an average accuracy of 93.5\%. For an additional cross-check, we also conducted an experiment where only a single test sample example was chosen to check the accuracy of this classifier. Based on a random one test sample of 106 known BL Lacs and 107 known FSRQS, our classifier returns 98 true FSRQs and 102 true BL Lacs. In other words, this  wrongly classified 4 BL Lacs and 9 FSRQs. This yielded an accuracy of $\sim$ 94\%, and is consistent with our more robust accuracy calculation described above. While this result might imply that the classifier is slightly biased towards finding the BL Lac class objects, this effect would be less than a few percent, based on these accuracy estimates.  Furthermore, we don't classify a source as a BL Lac or FSRQ when their respective probabilities are less than 90\%. Of course, while misclassifications are not expected, it is expected that the 'ambiguous/unclassified sources' could harbor both FSRQs and BL Lacs that could not be accurately classified.
\section{Results}\label{sec:results}
\begin{figure*}
    \centering
\includegraphics[width=\textwidth]{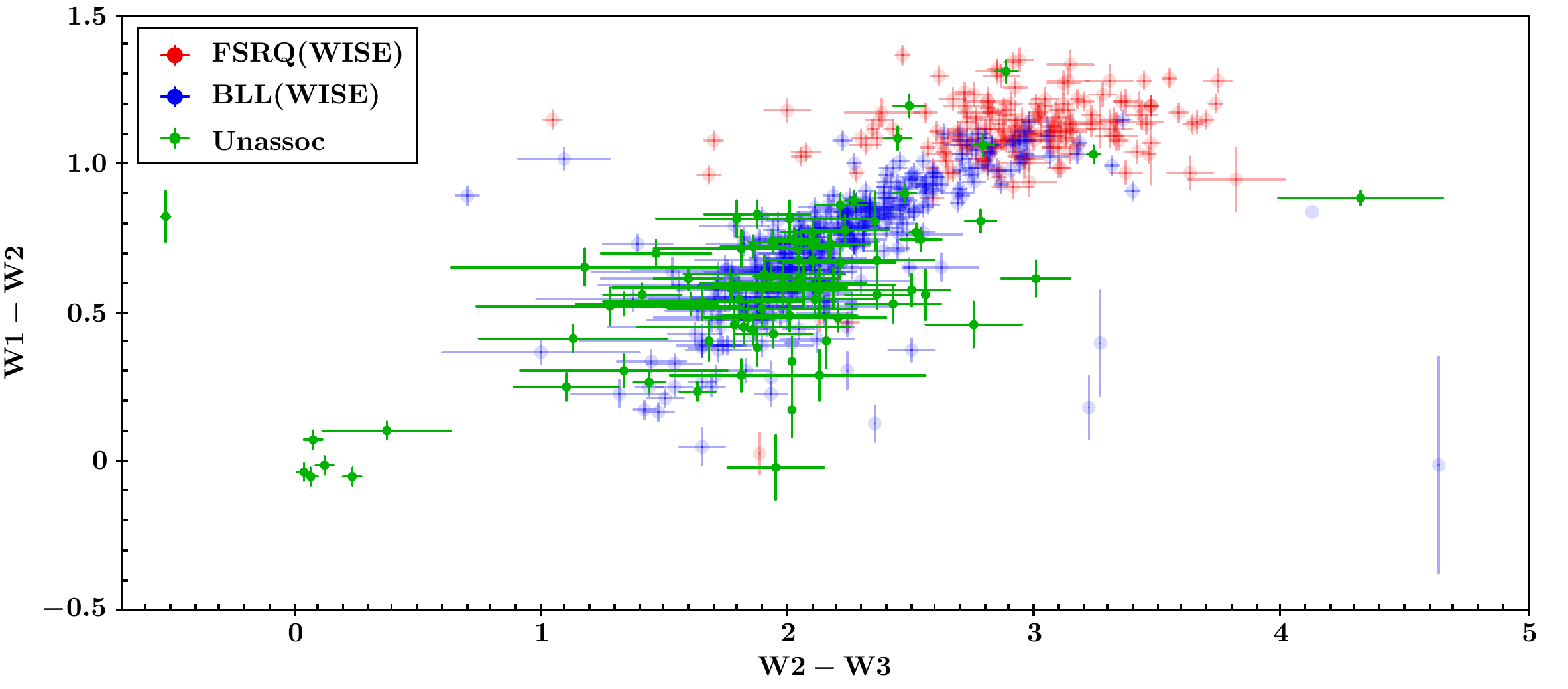}
    \caption{WISE blazar strip for the known \Fermi BL Lacs (blue) and FSRQs (red). Overplotted are the unassociated blazar candidates from this work (green). The W1, W2 and W3 correspond to the WISE filters; 3.4, 4.6 and 12 $\mu$ m, respectively. Please see Section~\ref{sec:Analysis} for complete details.}
    \label{fig:wise_all}
\end{figure*}

The trained classifier using the X-ray, gamma-ray and WISE parameters was then applied to the sample of 84 blazar candidate counterpart sources. Since the RF classifier provides a probability value for each source to belong to a particular class, we define the following classes based on the predicted probabilities:
BL Lac (bll): P$_{bll}$ $\ge$ 99\%, likely BL Lac: P$_{bll}$ $\ge$ 90\%, FSRQ: P$_{bll}$ $\le$ 1\%, likely FSRQ: P$_{bll}$ $\le$ 10\%, and ambiguous: 10\% $\ge$ P$_{bll}$ $\le$ 90\%; where P$_{bll}$ is the probability for a source to be a BL Lac. Using these definitions, we found that among the sample of 84 highly likely blazar candidates, 50 are likely BL Lacs and 34 are ambiguous. None of the sources were predicted to be FSRQs, nor did any of them fall into the category of likely FSRQs (See Table.~\ref{tab:results}). This is consistent with a visual inspection of Fig.~\ref{fig:wise_all}, which shows that our newly identified blazar candidate/counterpart sample seems to be constituted primarily of BL Lac class blazars, with a relatively small number of outliers that are ambiguously consistent with either the BL Lac or FSRQ classification and another small group of outliers that are ambiguous in the sense that they seem to fall outside of both the Bl Lac and the FSRQ distribution. The respective significances (percentage importances) of each parameter employed in the classifier are as follows:
X-ray flux: 0.058, Curvature: 0.062, Spectral Index: 0.241, Variability Index: 0.094, Gamma ray flux: 0.055, W1-W2: 0.312 and W2-W3: 0.175. 
\subsection{Miscellaneous-Outliers}\label{subsec:misc}
\begin{figure*}
    \centering
\includegraphics[width=\textwidth]{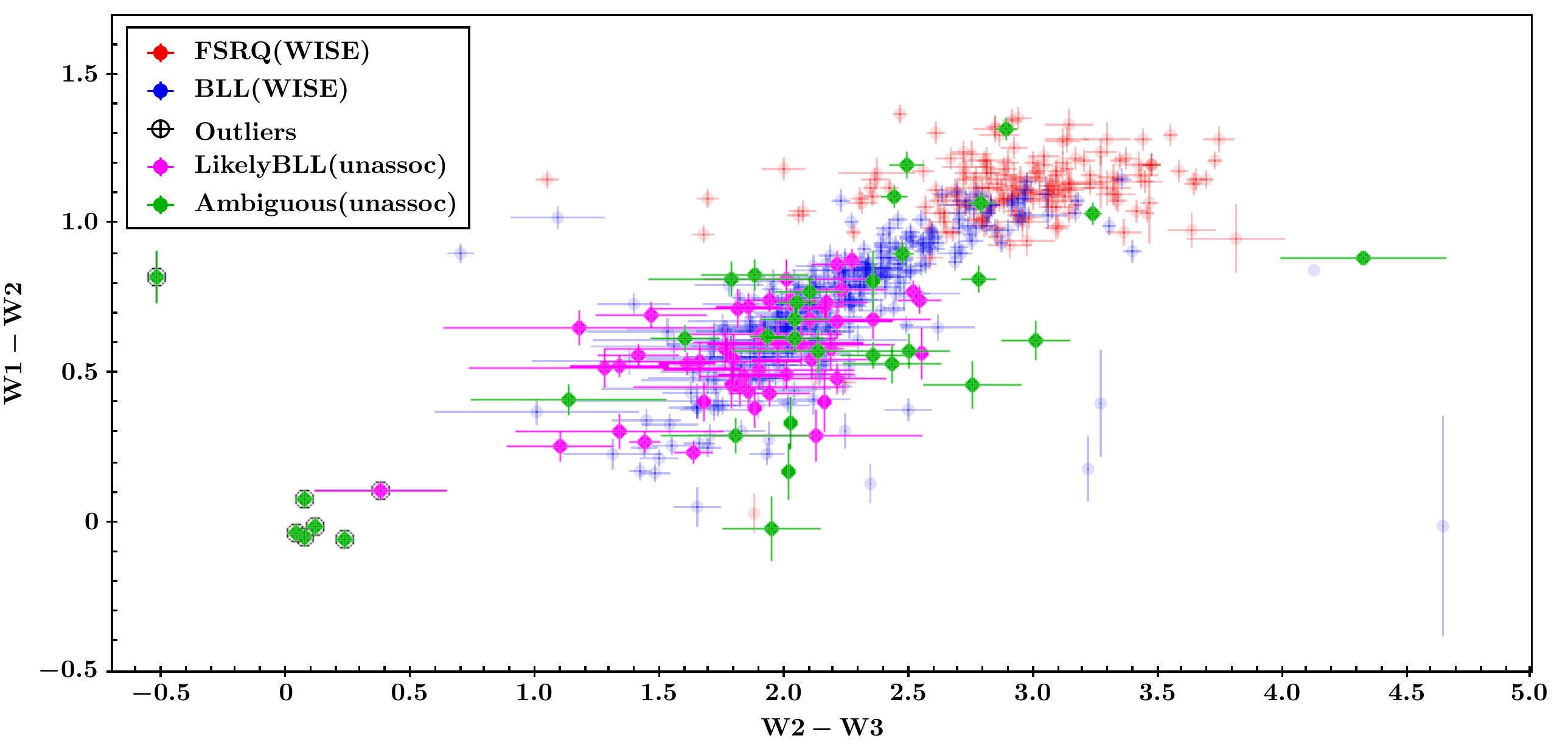}
    \caption{WISE blazar strip for the known \Fermi BL Lacs (blue) and FSRQs (red). Overplotted are the unassociated blazar candidates from this work as displayed in Fig.~\ref{fig:wise_all}. The subcategories displayed are based on the probabilities obtained with our machine learning algorithm dividing these into likely BL Lacs (magenta, P$_{bll}$ $\ge$ 90\%) and ambiguous blazars (green, 10\% $\ge$ P$_{bll}$ $\le$ 90\%). The outliers from the WBS are enclosed in $black$ circles. See the discussion in Section~\ref{subsec:misc} for a complete details on these sources. The W1, W2 and W3 correspond to the WISE filters; 3.4, 4.6 and 12 $\mu$m, respectively.}
    \label{fig:wise_subset}
\end{figure*}

\begin{figure*}
    \centering
\includegraphics[width=\textwidth]{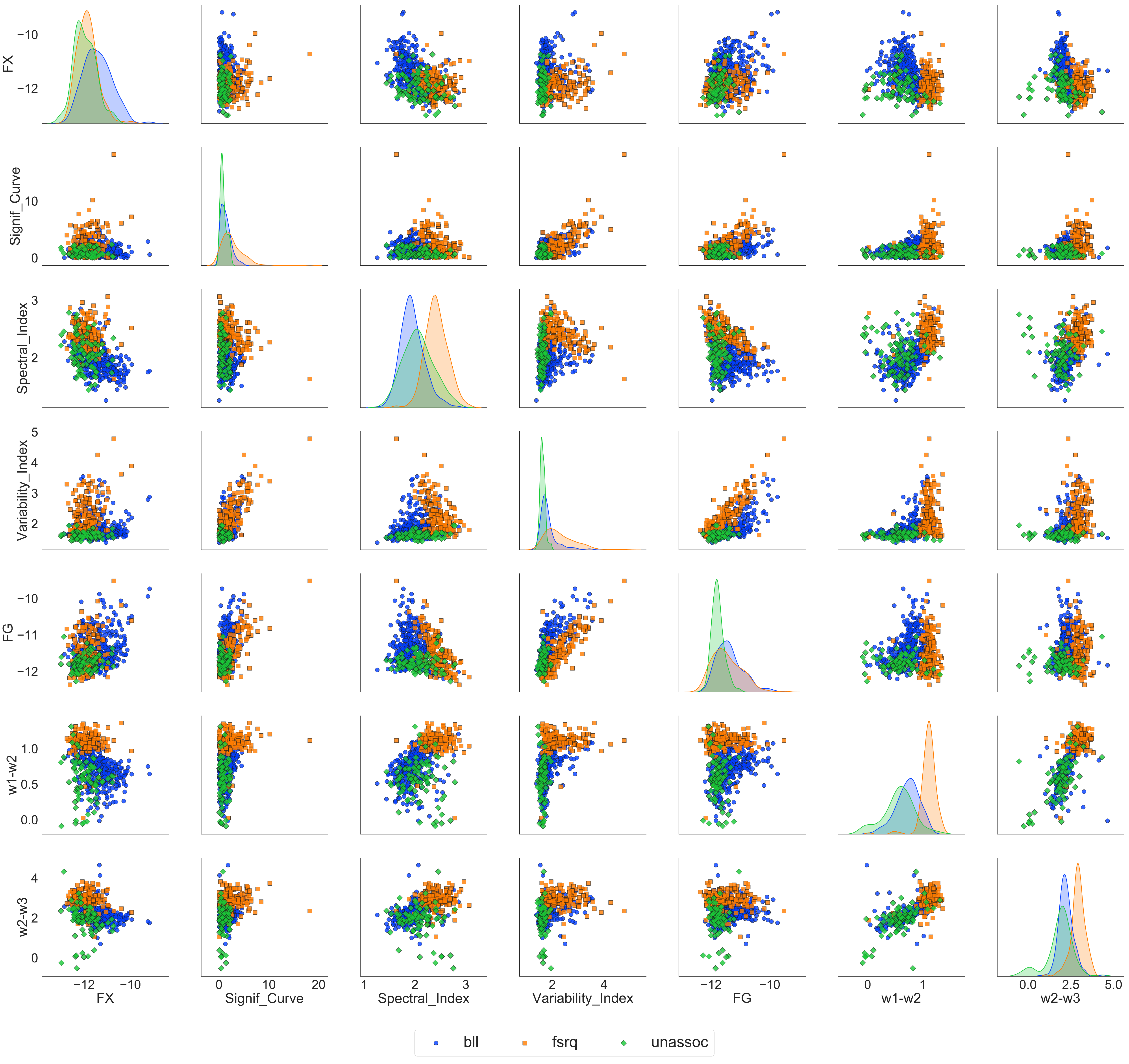}
    \caption{A comparison pairplot of seven parameters used to distinguish BL Lacs (bll; $blue$) from FSRQs (fsrq; $orange$). The 84 unassociated sources (unassoc; $green$) are also plotted. The plotted parameters are defined as follows: \texttt{FX, Signif\_Curve, Spectral\_Index, Variability\_Index, FG, w1-w2, w2-w3} represent log$_{10}$(X-ray Flux), Gamma-ray Curvature, Gamma-ray Spectral Index, log$_{10}$(Gamma-ray Variability Index), log$_{10}$(Gamma-ray Flux), WISE Color Index (W1-W2) and WISE Color Index (W2-W3), respectively.}
    \label{fig:pairplot}
\end{figure*}

A few of these unassociated source candidates seem to diverge from the usual WBS, as displayed in Fig.~\ref{fig:wise_subset} by using $black$ circular regions around them. These are seven in number out of which one belongs to the likely BL Lac and the rest to the ambiguous category. After further inspection, it was found that the positions of three of these were coincident with stars within 5 arcsec positional uncertainties of Swift$-$XRT positions. TYC4199-1248, which was also reported in K19, corresponds to a positional coincidence with 3FGL J1729.0+6049. Similarly, 3FGL J0748.8-2208 is spatially coincident with a star, TYX 5993-3722-1 and 3FGL JJ1801.5-7825 with HD 162298. No further information about these stars was found in literature. It is clear that the WISE position match yielded the colors for these non-blazar sources, which would explain their placement far left from the WBS in Fig.~\ref{fig:wise_subset}.  This is consistent with the fact that our ML method classified them as "ambiguous." In these few cases, it is also possible that the stellar systems could be associated with the source of gamma rays from the corresponding Fermi source. However, one of these outliers, namely 3FGL J1958.1+2436, is a confirmed BL Lac despite its position on the WISE color-color plot. Another interesting case is that of 3FGL J2035.8+4902, for which the position of the only X-ray source in the 3FGL error circle is positionally coincident with an eclipsing binary, V* V2552 Cyg. Some of these outliers may not belong to the blazar population and/or may not be the actual X-ray counterpart to the Fermi unassociated source, and various direct methods such as optical spectroscopy could be used to verify their true nature.  However, in some of the outlier cases, it is also possible that the detected X-ray source is a counterpart blazar which deviates from the usual gamma-ray blazar population and therefore should be further investigated for its interesting behavior. It should be noted that although the position of the few outliers on the WISE color-color plot does not mimic the other gamma-ray blazars, these parameters lie within the limits of the more general blazar population, particularly as an extension of the BL Lac distribution; please see Fig.~1 in \citet{Massaro2011}. Since the latest Fermi catalog Data Release 2 \citep[][4FGL-DR2]{Lott2020} was published recently, we compared our list of 84 sources to the source classifications in this release, for the cases where they are available. We found that 52 of the 84 sources are identified as bcus, and 7 are identified as BL Lacs. All except one (3FGL J0427.9-6704 classified as a "Binary" in the 4FGL catalog) of our results match with the 4FGL predictions as seen in Table~\ref{tab:results}, although three identified BL Lacs in the 4FGL catalog were characterized as "ambiguous" by our RF classifier (note: our classifier did find that these 3 sources were $>$ 81\% likely to be BL Lacs). In addition, we compared our classification results with a similar study conducted by \citet{Marchesini2020}, in which the authors searched for BL Lac candidates among the Fermi Unassociated sources. These authors selected their sample using a slightly different criteria such as SNR$>$3 as compared to ours with SNR$>$4. Moreover, these authors selected X-ray counterparts where more than one X-ray source was found within the \Fermi uncertainty region. Regardless, these authors found 19 sources in which are highly likely BL Lacs. Among these 19, we found 7 which are also present in our sample. 4 out of these 7 are identified as BL Lacs in our classification, whereas 3 are identified as ambiguous. Furthermore, it should be noted that two of these ambiguous sources have blazar probability $>$89\% in our classification, which makes them consistent with the BL Lac classification. Therefore, we consider our results to be consistent with this independent study conducted by these authors on this subset of the sources in our study.

\section{Conclusions}\label{sec:conclusions}
The immediate objective of this work is to classify, as either FSRQ or BL Lac, a sample of 84 highly likely X-ray blazar candidates that were drawn from a list of X-ray counterparts to 3FGL unassociated sources. This is a step forward towards the completeness of finding the gamma-ray emitting blazars. Finding associations to the unassociated gamma-ray emitting sources has been a necessary step in order to understand the gamma-ray emitting population in the Universe. Most of the gamma-ray sources belong to the category of blazars.  Classifying these blazars is an important step towards understanding their evolution and their role in galaxy evolution. Understanding the distribution of the subclasses for these gamma-ray emitting blazars plays a vital role in putting constraints on the blazar sequence \citep{Fossati1998,Ghisellini2017}. In previous work, we contributed to this task by finding counterparts and classifying the blazars among these unassociated sources, while this paper focuses on sub-classification of these blazar sources. Using the methods of machine learning, we find that 50 out of these 84 sources are $\ge$ 90\% likely to be BL Lacs, and the other 34 are not able to categorized (i.e. we categorize them as 'ambiguous'). However, since all these outliers/ambiguous sources are a subset of our blazar sample, these could be considered "bcus" (blazars of uncertain type). This implies that these sources are mostly likely either peculiar BL Lacs, FSRQs, or transitional blazars. Various follow up multiwavelength campaigns would be required to discern their nature. We don't find any of these sources to be clearly labeled as FSRQs. There could be multiple reasons for the paucity of sources categorized as FSRQs; e.g. (i) most of the X-ray counterparts to Fermi unassociated sources are indeed BL Lacs, which could be caused by inherent selection biases such as the fact that BL Lacs are more likely to have a synchrotron peak in the UV to X-ray band, (ii) The blazar component of Fermi unassociated sources has a selection bias that makes it more likely for an unassociated source to be a BL Lac, or, (iii) Since FSRQs are more bright and often have spectra available via various surveys, it is highly likely that most of the unassociated blazar population in the Fermi catalog are indeed BL Lacs. This pattern has also been seen in various optical spectroscopic surveys of unassociated Fermi sources, e.g., \citet{Crespo2016,Crespo2016b,Pena-Herazo2017,Paiano2017}
In the future, optical spectroscopic techniques can be utilized to find the nature of the 24 ambiguous sources, and to further investigate the properties of the classified blazars. Our study provides likely blazar targets for these spectroscopic optical observations, providing another avenue for localizing and characterizing possible counterparts with high precision.The ongoing questions regarding the Fermi blazar sequence and Fermi blazar divide require the redshift estimates for blazars. One should be able to confirm and determine the redshifts for the FSRQs found within the 24 sources by using various 4m class optical facilities, e.g. \citet{Crespo2016, Crespo2016b}. For the case of BL Lacs, traditionally 8-10m class telescopes \citep{Shaw2009, Shaw2013, Paiano2019} are utilized. This method of estimating the redshifts for BL Lacs is highly effective, but it is time and cost consuming. Recently, \citet{Rau2012} devised a photometric method to determine the redshifts (z), or find an upper limit, for BL Lacs.  One caveat of this method is that it works for sources with  z $>$ 1.3. This method has successfully found redshift estimates for 29 sources to-date \citep{Kaur2017, Kaur2018, Rajagopal2020}. 
We emphasize that our results are consistent with the updated 4FGL catalog classification as seen in Table.~\ref{tab:results}, for the newly classified sources, which provides further evidence that our added classifications (not classified in the 4FGL) are likely robust. In addition, we find three sources coincident with positions of known stars, as well as another source, 3FGL J2035.8+4902, spatially coincident with an eclipsing binary, V* V2552 Cyg. It should be noted that the latter is listed as a ''bcu'' in the 4FGL catalog based on a nearby fainter source at R.A.(J2000) : 20 35 51.63 and Dec(J2000) : +49 01 44.28, for which no WISE association could be found within a 5'' radius.  The further investigation of these sources is currently out of the scope of the work presented here, hence no further investigation was performed in this paper. 
\acknowledgments
The authors would like to gratefully acknowledge the support provided 
by NASA research grants \textcolor{blue}{80NSSC18K1730} and \textcolor{blue}{80NSSC19K1713}. M.C.S. is partially supported by the Heising-Simons Foundation under grant \#2018-0911. This publication made use of data products from the Wide- field Infrared Survey Explorer, which is a joint project of the University of California, Los Angeles, and the Jet Propulsion Laboratory/California Institute of Technology, funded by the National Aeronautics and Space Administration. The astronomical tool to compare databases, Topcat \citep{Taylor2005} was employed in this work. 

\clearpage

\startlongtable
\begin{deluxetable*}{CCcCl}
\tablecaption{Additional sample of blazar and pulsar candidates from the 3FGL unassociated sources since \citet{Kaur2019b} \label{tab:new sources}}
\tabletypesize{\scriptsize}
\tablecolumns{4}
\tablewidth{\columnwidth}
\tablehead{
\colhead{Swift Name\tablenotemark{a}} & \colhead{3FGL Name\tablenotemark{b}} &\colhead{Class\tablenotemark{c}} &
 \colhead{Random Forest\tablenotemark{d}} \\
\colhead{SwF3}  &\colhead{3FGL} &  &
 \colhead{BLL Probability}} 
\startdata
  J024454.9+475117 & J0244.4+4745 & ambiguous &0.28  \\
  J034158.1+314851 & J0342.3+3148c & ambiguous &0.455  \\
  J053559.3-061624 & J0535.7-0617c & blazar  &0.997  \\
  J071046.2-102942 & J0711.1-1037 & ambiguous & 0.153  \\
  J090530.4-491840 & J0905.6-4917 & likely pulsar &0.048 \\
  J091926.1-220043 & J0919.5-2200 & blazar &0.994  \\
  J101545.9-602938 & J1016.5-6034 & ambiguous &0.166  \\
  J103831.1-581346 & J1039.1-5809 & ambiguous &0.149  \\
  J105040.6-611608 & J1050.6-6112 & ambiguous &0.407  \\
  J105224.5+081409 & J1052.0+0816 & blazar &0.996  \\
  J110025.5-205333 & J1100.2-2044 & blazar &1.0  \\
  J110224.1-773339 & J1104.3-7736c & ambiguous &0.438  \\
  J111601.8-484222 & J1116.7-4854 & ambiguous &0.624  \\
  J121553.0-060940 & J1216.6-0557 & likely blazar &0.954  \\
  J130128.9+333711 & J1301.5+3333 & ambiguous&0.283  \\
  J132928.6-053135 & J1329.1-0536 & blazar &1.0  \\
  J162437.8-423144 & J1624.8-4233 & ambiguous&0.892  \\
  J162743.0+322102 & J1627.8+3217 &ambiguous &0.279  \\
  J170521.6-413436 & J1705.5-4128c &ambiguous &0.484  \\
  J174511.10-225455 & J1744.7-2252 &ambiguous &0.122  \\
  J185520.0+075138 & J1855.6+0753 &ambiguous &0.737  \\
  J201020.3-212434 & J2010.0-2120 & ambiguous&0.772  \\
  J205350.8+292312 & J2053.9+2922 &ambiguous &0.797  \\
  J212051.6-125300 & J2120.4-1256 &ambiguous &0.884  \\
  J232653.3-412713 & J2327.2-4130 & blazar &0.998  \\
  \enddata
  \tablenotetext{a}{The name of the \Swift discovered X-ray source within the 95\% \Fermi uncertainty region of the corresponding 3FGL source as defined in K19.}
  \tablenotetext{b}{The Fermi source name as defined in 3FGL catalog.}
  \tablenotetext{c}{The Classification based on the probability of a given source to be called a blazar/pulsar/ambiguous as defined in K19 and also described in Section~\ref{sec:sample}}
  \tablenotetext{d}{The probability of a given source to be identified a BL Lac, derived from the random forest classifier.}
\end{deluxetable*}
\startlongtable
\begin{deluxetable*}{CCccCl}
\tablecaption{Classification using Machine Learning : BLLs and FSRQs  \label{tab:results}}
\tabletypesize{\scriptsize}
\tablecolumns{6}
\tablewidth{\columnwidth}
\tablehead{
\colhead{Swift Name\tablenotemark{a}} & \colhead{3FGL Name\tablenotemark{b}} & \colhead{Expected \# of\tablenotemark{c}} & \colhead{Class\tablenotemark{d}} &
 \colhead{Random Forest\tablenotemark{e}} & \colhead{Notes}\\
\colhead{SwF3}  &\colhead{3FGL} & \colhead{chance X-ray sources} & &
 \colhead{BLL Probability}& \colhead{Classification in literature}} 
\startdata
J004859.5+422348&J0049.0+4224&0.00&likely bll&0.955&bcu\citep[4FGL,][]{Lott2020}\\
J015624.4-242003&J0156.5-2423&0.50&likely bll&0.957&\\
J020020.9-410933&J0200.3-4108&0.00&likely bll&0.957&bcu\citep[4FGL,][]{Lott2020}\\
J023854.1+255405&J0239.0+2555&0.04&likely bll&0.93&bcu\citep[4FGL,][]{Lott2020}\\
J025857.4+055243&J0258.9+0552&0.04&ambiguous&0.855&\\
J034050.0-242259&J0340.4-2423&0.03&ambiguous&0.512&bcu\citep[4FGL,][]{Lott2020}\\
J034819.8+603506&J0348.4+6039&0.04&likely bll&0.954&bcu\citep[4FGL,][]{Lott2020}\\
J035051.2-281632&J0351.0-2816&0.33&likely bll&0.912&bcu\citep[4FGL,][]{Lott2020}\\
J042011.0-601504&J0420.4-6013&0.07&likely bll&0.958&bcu\citep[4FGL,][]{Lott2020}\\
J042749.8-670434&J0427.9-6704&0.40&ambiguous&0.805&LMXB \citep[4FGL,][]{Lott2020}\\
J042958.7-305931&J0430.1-3103&0.00&likely bll&0.939&bcu\citep[4FGL,][]{Lott2020}\\
J043836.9-732919&J0437.7-7330&0.00&likely bll&0.934&bcu\citep[4FGL,][]{Lott2020}\\
J050650.1+032359&J0506.9+0321&0.60&likely bll&0.958&bcu\citep[4FGL,][]{Lott2020}\\
J051641.4+101243&J0516.6+1012&0.15&likely bll&0.942&\\
J074626.1-022551&J0746.4-0225&0.01&likely bll&0.956&bcu\citep[4FGL,][]{Lott2020}\\
J074724.8-492633&J0747.5-4927&0.01&ambiguous&0.898&bcu\citep[4FGL,][]{Lott2020}\\
J074903.8-221015&J0748.8-2208&0.00&ambiguous&0.747&\\
J080215.8-094214&J0802.3-0941&0.01&ambiguous&0.894&bcu\citep[4FGL,][]{Lott2020}\\
J081338.1-035717&J0813.5-0356&0.00&likely bll&0.949&bcu\citep[4FGL,][]{Lott2020}\\
J082628.2-640415&J0826.3-6400&0.07&ambiguous&0.871&bll \citep[4FGL, ][]{Lott2020}\\
J084831.8-694108&J0847.2-6936&0.00&ambiguous&0.818&\\
J091926.1-220043&J0919.5-2200&1.85&ambiguous&0.888&bcu\citep[4FGL,][]{Lott2020}\\
J092818.4-525659&J0928.3-5255&0.00&likely bll&0.931&\\
J093754.6-143349&J0937.9-1435&0.00&likely bll&0.954&bcu\citep[4FGL,][]{Lott2020}\\
J095249.5+071329&J0952.8+0711&0.00&likely bll&0.953&bcu\citep[4FGL,][]{Lott2020}\\
J102432.6-454428&J1024.4-4545&0.00&likely bll&0.948&\\
J103332.4-503526&J1033.4-5035&0.00&likely bll&0.957&bcu\citep[4FGL,][]{Lott2020}\\
J105224.5+081409&J1052.0+0816&0.00&ambiguous&0.882&bcu\citep[4FGL,][]{Lott2020}\\
J110025.5-205333&J1100.2-2044&0.26&ambiguous&0.695&\\
J113032.7-780107&J1130.7-7800&0.00&ambiguous&0.899&bcu\citep[4FGL,][]{Lott2020}\\
J113209.3-473853&J1132.0-4736&0.10&likely bll&0.937&bcu\citep[4FGL,][]{Lott2020}\\
J120055.1-143039&J1200.9-1432&0.00&likely bll&0.906&bcu\citep[4FGL,][]{Lott2020}\\
J122014.4-245948&J1220.0-2502&0.03&likely bll&0.904&bcu\citep[4FGL,][]{Lott2020}\\
J122019.8-371414&J1220.1-3715&0.00&likely bll&0.902&\\
J122127.4-062845&J1221.5-0632&0.23&ambiguous&0.82&\\
J122257.0+121438&J1223.2+1215&2.72&ambiguous&0.847&bll\citep[4FGL,][]{Lott2020}\\
J122536.7-344723&J1225.4-3448&0.00&likely bll&0.955&bcu\citep[4FGL,][]{Lott2020}\\
J123235.9-372055&J1232.5-3720&0.00&likely bll&0.955&bcu\citep[4FGL,][]{Lott2020}\\
J123726.6-705140&J1236.6-7050&0.00&ambiguous&0.521&\\
J124021.3-714858&J1240.3-7149&0.02&likely bll&0.955&bcu\citep[4FGL,][]{Lott2020}\\
J124919.5-280833&J1249.1-2808&0.18&likely bll&0.946&bcu\citep[4FGL,][]{Lott2020}\\
J124919.7-054540&J1249.5-0546&0.08&ambiguous&0.755&bcu\citep[4FGL,][]{Lott2020}\\
J131140.3-623313&J1311.8-6230&1.63&ambiguous&0.657&\\
J131552.8-073304&J1315.7-0732&0.00&likely bll&0.952&bcu\citep[4FGL,][]{Lott2020}\\
J141133.3-072256&J1411.4-0724&0.00&likely bll&0.955&bcu\citep[4FGL,][]{Lott2020}\\
J151150.1+662450&J1512.3+6622&0.00&likely bll&0.906&\\
J151649.8+263635&J1517.0+2637&0.00&ambiguous&0.501&bcu\citep[4FGL,][]{Lott2020}\\
J152603.0-083146&J1525.8-0834&0.00&likely bll&0.946&bcu\citep[4FGL,][]{Lott2020}\\
J152818.2-290256&J1528.1-2904&0.02&ambiguous&0.833&bcu\citep[4FGL,][]{Lott2020}\\
J154150.1+141441&J1541.6+1414&0.00&likely bll&0.955&bll\citep[4FGL,][]{Lott2020}\\
J154946.4-304502&J1549.9-3044&0.00&likely bll&0.903&bcu\citep[4FGL,][]{Lott2020}\\
J170409.6+123423&J1704.1+1234&0.00&ambiguous&0.837&bll\citep[4FGL,][]{Lott2020}\\
J171106.1-432415&J1710.6-4317&0.00&ambiguous&0.774&\\
J172142.1-392204&J1721.8-3919&0.20&ambiguous&0.752&\\
J172858.2+604359&J1729.0+6049&0.00&ambiguous&0.751&\\
J180106.8-782248&J1801.5-7825&0.00&ambiguous&0.697&\\
J192242.1-745354&J1923.2-7452&0.00&likely bll&0.901&bcu\citep[4FGL,][]{Lott2020}\\
J193420.1+600138&J1934.2+6002&0.00&ambiguous&0.897&bcu\citep[4FGL,][]{Lott2020}\\
J195149.7+690719&J1951.3+6909&0.00&ambiguous&0.721&\\
J195800.3+243803&J1958.1+2436&0.00&likely bll&0.941&bcu\citep[4FGL,][]{Lott2020}\\
J201525.3-143204&J2015.3-1431&0.00&likely bll&0.956&bll\citep[4FGL,][]{Lott2020}\\
J203027.9-143919&J2030.5-1439&0.00&likely bll&0.951&bcu\citep[4FGL,][]{Lott2020}\\
J203450.9-420038&J2034.6-4202&0.01&likely bll&0.946&bcu\citep[4FGL,][]{Lott2020}\\
J203556.9+490038$^\star$&J2035.8+4902&0.00&ambiguous&0.877&bcu\citep[4FGL,][]{Lott2020}\\
J203935.8+123001&J2039.7+1237&0.00&ambiguous&0.59&\\
J205950.4+202905&J2059.9+2029&0.00&ambiguous&0.826&\\
J210940.0+043958&J2110.0+0442&0.00&likely bll&0.931&bll\citep[4FGL,][]{Lott2020}\\
J211522.2+121801&J2115.2+1215&0.00&likely bll&0.917&bcu\citep[4FGL,][]{Lott2020}\\
J214247.5+195811&J2142.7+1957&0.01&likely bll&0.936&bcu\citep[4FGL,][]{Lott2020}\\
J215122.1+415634&J2151.6+4154&0.00&likely bll&0.949&bll\citep[4FGL,][]{Lott2020}\\
J222911.2+225456&J2229.1+2255&0.00&likely bll&0.912&bcu\citep[4FGL,][]{Lott2020}\\
J224437.0+250344&J2244.6+2503&0.00&ambiguous&0.85&bcu\citep[4FGL,][]{Lott2020}\\
J225032.7+174918&J2250.3+1747&0.08&ambiguous&0.702&bcu\citep[4FGL,][]{Lott2020}\\
J230012.4+405223&J2300.0+4053&0.27&likely bll&0.949&bll\citep[4FGL,][]{Lott2020}\\
J230351.7+555617&J2303.7+5555&0.00&likely bll&0.924&\\
J230848.5+542612&J2309.0+5428&0.17&likely bll&0.956&\\
J232127.1+511117&J2321.3+5113&0.00&ambiguous&0.887&bcu\citep[4FGL,][]{Lott2020}\\
J232137.1-161926&J2321.6-1619&0.00&likely bll&0.925&bcu\citep[4FGL,][]{Lott2020}\\
J232653.3-412713&J2327.2-4130&0.00&ambiguous&0.46&\\
J232938.7+610111&J2329.8+6102&0.00&likely bll&0.931&bcu\citep[4FGL,][]{Lott2020}\\
J233626.4-842649&J2337.2-8425&0.01&ambiguous&0.837&\\
J235115.9-760017&J2351.9-7601&0.00&likely bll&0.959&bll\citep[4FGL,][]{Lott2020}\\
J235824.1+382857&J2358.5+3827&0.00&likely bll&0.954&bll\citep[4FGL,][]{Lott2020}\\
J235836.8-180717&J2358.6-1809&0.10&likely bll&0.936&bcu\citep[4FGL,][]{Lott2020}\\
\enddata
  \tablenotetext{a}{The name of the \Swift discovered X-ray source within the 95\% \Fermi uncertainty region of the corresponding 3FGL source as defined in K19.}
  \tablenotetext{b}{The Fermi source name as defined in 3FGL catalog.}
  \tablenotetext{c}{The expected number of X-ray sources found within the Fermi uncertainty ellipse using \Swift-XRT. See section~\ref{sec:sample} for further details}
  \tablenotetext{d}{The Classification based on the probability of a given source to be called a BL Lac/FSRQ/ambiguous as defined in this work. See Section~\ref{sec:results}.}
  \tablenotetext{e}{The probability of a given source to be identified a BL Lac, derived from the random forest classifier.}
 \tablecomments{$\star$: The Swift X-ray position of this source is positionally coincident with an eclipsing binary, V* V2552 Cyg}
\end{deluxetable*}
\clearpage

\vspace{5mm}
\facilities{Swift(XRT), WISE}

\bibliography{bibliography}{}

\bibliographystyle{aasjournal}

\end{document}